\documentclass{article}

\PassOptionsToPackage{numbers,sort,compress}{natbib}

    \usepackage[final]{neurips_2024}

\usepackage[utf8]{inputenc} 
\usepackage[T1]{fontenc}    
\usepackage[breaklinks=true,colorlinks,bookmarks=True]{hyperref} 
\usepackage{url}            
\usepackage{booktabs}       
\usepackage{amsfonts}       
\usepackage{nicefrac}       
\usepackage{microtype}      
\usepackage{xcolor}         

\newcommand{\eat}[1]{} 
\newcommand{\blue}[1]{ {\textcolor{blue}{#1}}}

\newcommand{\clidth}{$\text{CLiD}_{th}$\xspace}
\newcommand{\clidvec}{$\text{CLiD}_{vec}$\xspace}
\newcommand{\vect}[1]{\mathbf{#1}}

\usepackage[framemethod=TikZ]{mdframed}
\usepackage{amsmath}
\usepackage{amsthm}
\usepackage{bbm}
\usepackage{subfigure}
\usepackage{fancyhdr}
\usepackage{xspace}
\usepackage{textcomp}
\usepackage{enumerate}
\usepackage{threeparttable}
\usepackage{soul}

\usepackage{listings}
\usepackage{multirow} 
\usepackage[linesnumbered,ruled]{algorithm2e}
\usepackage{mathtools}
\usepackage{wrapfig}
\usepackage{booktabs}
\usepackage{lipsum}
\makeatletter \renewcommand{\@thesubfigure}{\thesubfigure \space}%

\usepackage{tabularx} %

\theoremstyle{plain}
\newtheorem{theorem}{Theorem}[section]

\newtheorem{assumption}[theorem]{Assumption}
\theoremstyle{remark}

\usepackage[capitalize]{cleveref}
\crefname{section}{Sec.}{Secs.}
\Crefname{section}{Section}{Sections}
\Crefname{table}{Table}{Tables}
\crefname{table}{Tab.}{Tabs.}

\title{
Membership Inference on Text-to-image Diffusion Models via Conditional Likelihood Discrepancy
}

\author{
Shengfang Zhai$^{1,2}$,
Huanran Chen$^{3,6}$, 
Yinpeng Dong$^{3,6}$\thanks{Corresponding authors.}\;, 
Jiajun Li$^{1,2}$,
\and
\textbf{
Qingni Shen$^{1,2}$\footnotemark[1]\;, 
Yansong Gao$^{4}$,
Hang Su$^{3,5}$, Yang Liu$^{7}$} \\
$^{1}$School of Software and Microelectronics, Peking University 
\\ 
$^{2}$PKU-OCTA Laboratory for Blockchain and Privacy Computing, Peking University \\
$^{3}$Dept. of Comp. Sci. and Tech., Institute for AI, BNRist Center, THBI Lab, Tsinghua University\\
$^{4}$The University of Western Australia \quad $^5$Zhongguancun Laboratory, Beijing, China \\
$^{6}$RealAI \quad 
$^{7}$Nanyang Technological University \\
\texttt{\{zhaisf,  jiajun.lee\}@stu.pku.edu.cn} \quad  
\texttt{huanran.chen@outlook.com} \\  
\texttt{\{dongyinpeng, suhangss\}@tsinghua.edu.cn} \\
\texttt{qingnishen@ss.pku.edu.cn} \quad
\texttt{garrison.gao@uwa.edu.au} \\  
\texttt{yangliu@ntu.edu.sg} 
}

\begin{document}

\maketitle

\begin{abstract}
Text-to-image diffusion models have achieved tremendous success in the field of controllable image generation, while also coming along with issues of privacy leakage and data copyrights. Membership inference arises in these contexts as a potential auditing method for detecting unauthorized data usage. While some efforts have been made on diffusion models, they are not applicable to text-to-image diffusion models due to the high computation overhead and enhanced generalization capabilities. In this paper, we first identify a conditional overfitting phenomenon in text-to-image diffusion models, indicating that these models tend to overfit the conditional distribution of images given the corresponding text rather than the marginal distribution of images only. 
Based on this observation, we derive an analytical indicator, namely \textbf{C}onditional \textbf{Li}kelihood \textbf{D}iscrepancy (\textbf{CLiD}), to perform membership inference, which reduces the stochasticity in estimating memorization of individual samples.
Experimental results demonstrate that our method significantly outperforms previous methods across various data distributions and dataset scales. Additionally, our method shows superior resistance to overfitting mitigation strategies, such as early stopping and data augmentation.
\end{abstract}

\section{Introduction}

Text-to-image diffusion models have achieved remarkable success in the guided generation of diverse, high-quality images based on text prompts, 
such as Stable Diffusion \cite{rombach2022high, podell2023sdxl}, DALLE-2 \cite{ramesh2022hierarchical}, Imagen \cite{saharia2022photorealistic}, and DeepFloyd-IF \cite{DeepFloyd_IF}. 
These models are increasingly adopted by users to create photorealistic images that align with desired semantics.
Moreover, they can generate images of specific concepts \cite{Pokemon-Blip}  or styles  \cite{Wikiart} when fine-tuned on relevant datasets.
However, the impressive generative capabilities of these models depend heavily on high-quality image-text datasets, which involve collecting image-text data from the web. 
This practice raises significant privacy and copyright concerns in the community~\cite{carlini2023extracting,Generative_AI_Has}. 
The pretraining and fine-tuning processes of text-to-image diffusion models can cause copyright infringement, as they utilize unauthorized datasets published by human artists or stock-image websites 
\cite{bbc2022Art, cnn2022AI, washington2022AI, Reuters2023getty, Reuters2023Lawsuits}. 

Membership inference (also known as the membership inference attack) is widely used for auditing privacy leakage of training data 
\cite{shokri2017membership, carlini2022membership}, defined as determining whether a given data point has been used to train the target model. 
Dataset owners can thus leverage membership inference to determine if their data is being used without authorization \cite{miao2021audio,dealcala2024my}.

Previous works \cite{carlini2023extracting,duan2023diffusion,kong2023efficient,fu2023probabilistic,dubinski2024towards,matsumoto2023membership} have attempted membership inference on diffusion models. 
\citet{carlini2023extracting} employ LiRA (Likelihood Ratio Attack)~\cite{carlini2022membership} to perform membership inference on diffusion models. 
LiRA requires training multiple shadow models to estimate the likelihood ratios of a data point from different models, which incurs high training overhead (e.g., 16 shadow models for DDPM \cite{ho2020denoising} on CIFAR-10 \cite{krizhevsky2009learning}), making it neither scalable nor applicable to text-to-image diffusion models. Other query-based membership inference methods \cite{matsumoto2023membership,duan2023diffusion,kong2023efficient,fu2023probabilistic} design and compute indicators to evaluate whether a given data point belongs to the member set. These methods require only a few or even a single shadow model, making them scalable to larger text-to-image diffusion models. 
However, these methods mainly estimate model memorization for data points and do not fully utilize the conditional distribution of image-text pairs.
Consequently, they achieve limited success only under excessively high training steps and fail under real steps or common data augmentation methods (Tab.~\ref{table:results_real}), which do not reflect real training scenarios. 
Text-to-image diffusion models have demonstrated excellent performance in zero-shot image generation \cite{rombach2022high, podell2023sdxl,bao2023one}, indicating their strong generalization, which makes it difficult to distinguish membership by directly measuring overfitting to data points.  
And due to the stochasticity of diffusion training loss~\cite{ho2020denoising, rombach2022high}, this kind of measuring becomes more challenging.


To address the challenges, we firstly identify a \textbf{Conditional Overfitting} phenomenon of text-to-image diffusion models with empirical validation, where the models exhibit more significant overfitting to the conditional distribution of the images given the corresponding text than the marginal distribution of the images only. It inspires the revealing of membership by leveraging the overfitting difference.
Based on it, we propose to perform membership inference on text-to-image diffusion models via \textbf{C}onditional \textbf{Li}kelihood \textbf{D}iscrepancy (\textbf{CLiD}).
Specifically, CLiD quantifies overfitting difference analytically by utilizing Kullback-Leibler (KL) divergence as the distance metric and derives a membership inference indicator that estimates the discrepancy between the conditional likelihood of image-text pairs and the likelihood of images only. 
We approximate the likelihoods by employing Monte Carlo sampling on their ELBOs (Evidence Lower Bounds),
and design two membership inference methods: a threshold-based method \clidth and a feature vector-based method \clidvec.

We conduct extensive experiments on three text-to-image datasets~\cite{lin2014microsoft,young2014image,Pokemon-Blip} with various data distributions and dataset scales, using the mainstream open-sourced text-to-image diffusion models \cite{Stable-Diffusion-v1-4, Stable-Diffusion-v1-5} under both fine-tuning and pretraining settings. 
First, our methods consistently outperform existing baselines across various data distributions and training scenarios, including fine-tuning settings and the pretraining setting. Second, our experiments on fine-tuning settings with different training steps (Sec.~\ref{sec:main_result}) reveal that excessively high step/image ratios cause overfitting, leading to hallucination success; and we develop a more realistic pretraining setting following \cite{dubinski2024towards,das2024blind}, where our experiments reveal the insufficient effect of existing membership inference works \cite{duan2023diffusion, kong2023efficient, fu2023probabilistic, matsumoto2023membership}.
Third, our comparison experiment with varying training steps (Sec.~\ref{sec:vary_steps}) indicates that the effectiveness of membership inference grows with higher step/image ratios and should be evaluated under reasonable settings for realistic results.
Next, ablation studies further demonstrate the effect of our CLiD indicator, even with fewer query count, our method still outperforms baseline methods (\cref{fig:auc_samples}).
Last, experiments show that our methods exhibit stronger resistance to data augmentation, and exhibit resistance to even adaptive defenses.

\vspace{-2pt}
\section{Diffusion Model Preliminaries}
\vspace{-1pt}

\textbf{Denoising Diffusion Probabilistic Model (DDPM)} \cite{ho2020denoising} learns the data distribution 
$\mathbf{x}_0 \sim q(\mathbf{x}) $ 
by reversing the forward noise-adding process. 
For the forward process, DDPM defines a Markov process of adding Gaussian noise step by step:
\begin{equation}
    q(\mathbf{x}_t | \mathbf{x}_{t-1}) := \mathcal{N}(\mathbf{x}_t; \sqrt{1 - \beta_t} \mathbf{x}_{t-1}, \beta_t \mathbf{I}),
    \label{eq:ddpm_forward}
\end{equation}
where $\beta_{t} \in (0, 1)$ is the hyperparameter controlling the variance. 
For the reverse process, DDPM defines a learnable Markov chain starting at $p(\mathbf{x}_T)=\mathcal{N}(\mathbf{x}_T;\mathbf{0},\mathbf{I})$ to generate $\mathbf{x}_0$:
\begin{equation}\label{eq:reverse}
p_\theta(\mathbf{x}_{0}) =  \int_{\mathbf{x}_{1:T}} p(\mathbf{x}_T)\prod_{t=1}^T p_\theta(\mathbf{x}_{t-1}|\mathbf{x}_t)\  \mathrm{d}\mathbf{x}_{1:T}, 
\qquad 
    p_{\theta}(\mathbf{x}_{t-1} | \mathbf{x}_{t}) = \mathcal{N}(\mathbf{x}_{t-1}; \mathbf{\mu}_{\theta}(\mathbf{x}_{t}, t), \sigma_t^2),
\end{equation}
where $\sigma_t^2$ is the untrained time-dependent constant.\,$\theta$ represents the trainable parameters.
To maximize $p_\theta(\mathbf{x}_0)$, DDPM optimizes the Evidence Lower Bound (ELBO) of the log-likelihood~\cite{ho2020denoising, li2023your}:
\begin{equation}\label{eq:ddpm_elbo}
       \log p_\theta (\mathbf{x}_0 )  
       \geq \mathbb{E}_{q(\mathbf{x}_{1:T} \mid \mathbf{x}_0)}\left[\log \frac{p_\theta(\mathbf{x}_{0:T})}{q(\mathbf{x}_{1:T} \mid \mathbf{x}_0)}\right] 
   = - \mathbb{E}_{ \mathbf{\epsilon}, t} \left[|| \mathbf{\epsilon}_{\theta}( \mathbf{x}_t    , t) - \epsilon ||^2 \right] + C,
\end{equation}
where $\epsilon \sim \mathcal{N}(0,\mathcal{I})$,  $t \sim \text{Uniform}({1,...,T}) $ and $C$ is a constant. $\mathbf{x}_t$ 
is obtained from Eq.~\eqref{eq:ddpm_forward},
and $\epsilon_\theta$ is 
a function approximator intended to predict the noise $\epsilon$ from $\mathbf{x}_t$. Omitting the untrainable constant in Eq.~\eqref{eq:ddpm_elbo} and taking its negative yields the loss function of training DDPM.

\textbf{Conditional diffusion models} \cite{ho2022classifier, nichol2021glide, rombach2022high}.
To achieve controllable generation ability, text-to-image diffusion models
incorporate the conditioning mechanism into the model, which are also known as conditional diffusion models, enabling them to learn conditional  probability as:
\begin{equation}
p_\theta(\mathbf{x}_0 |\mathbf{c}) = \int_{\mathbf{x}_{1:T}} p(\mathbf{x}_T) \prod_{t=1}^T p_\theta(\mathbf{x}_{t-1} | \mathbf{x}_t, \mathbf{c})\  \mathrm{d}\mathbf{x}_{1:T},
\end{equation}
where $\mathbf{c}$ denotes the embedding of condition. For text-to-image synthesis, $\mathbf{c}:=\mathcal{T}(\mathbf{y})$, where  $\mathbf{y}$ and $\mathcal{T}$ denote the text input and the text encoder, respectively.
Similar to Eq.~\eqref{eq:ddpm_elbo}, through derivation \cite{li2023your}, we can obtain the ELBO of the conditional log-likelihood:
\begin{equation}\label{eq:cdm_elbo}
   \log p_\theta(\mathbf{x}_0 |\mathbf{c}) 
   \geq
   - \mathbb{E}_{ \mathbf{\epsilon}, t} \left[|| \mathbf{\epsilon}_{\theta}( \mathbf{x}_t    , t, \mathbf{c}) - \epsilon ||^2 \right] + C.
\end{equation}

\section{Methodology} 

In this section, we detail the proposed \textbf{C}onditional \textbf{Li}kelihood \textbf{D}iscrepancy (\textbf{CLiD}) method. We first introduce the threat model of query-based membership inference in \cref{sec:threat_m}. We then identify the conditional overfitting phenomenon with experimental validation in \cref{sec:key_i}. We further drive the membership inference indicator based on CLiD in \cref{sec:clid} and design two practical membership inference methods in \cref{sec:clid_mi}.
We finally provide the implementation details in \cref{sec:prac}.

\vspace{-2pt}
\subsection{Threat Model}\label{sec:threat_m}
\vspace{-2pt}

We use the standard security game of membership inference on image-text data following previous work \cite{sablayrolles2019white, carlini2022membership, carlini2023extracting, matsumoto2023membership}.
We define a challenger $\mathcal{C}$  and an adversary $\mathcal{A}$ who performs membership inference. $\mathcal{C}$ samples a member set $D_{\text{mem}} \leftarrow \mathbb{D}$ and trains or fine-tunes a text-to-image diffusion model $f_{\theta}$ (i.e., target model) with $D_{\text{mem}}$. The rest of $\mathbb{D}$ is denoted by hold-out set $D_{\text{out}}$ =  $\mathbb{D} \setminus D_{\text{mem}}$. 
For a given data point ${(\mathbf{x}, \mathbf{c}}) \in \mathbb{D} $, $\mathcal{A}$ designs an algorithm $\mathcal{M}$ to yield a membership prediction:
\begin{equation}  
\mathcal{M}(\mathbf{x},\mathbf{c},f_{\theta}) = \mathbbm{1} \left[
\mathcal{M'}(\mathbf{x},\mathbf{c},f_{\theta})
> \tau  \right],
\end{equation}
where $\mathcal{M'}$ denotes an indicator function that reflects membership information, and $\tau $ denotes a tunable decision threshold of query-based membership inference \cite{matsumoto2023membership,duan2023diffusion, kong2023efficient,fu2023probabilistic}.

We consider a grey-box setting
\footnote{Note that in most real-world scenarios, the requirements for  $\mathcal{A}$ in gray-box and white-box settings are nearly identical. We use this terminology here for consistency with previous works~\cite{duan2023diffusion, kong2023efficient}.} 
consistent with previous query-based methods \cite{matsumoto2023membership,duan2023diffusion, kong2023efficient, fu2023probabilistic}. This setting assumes that  $\mathcal{A}$ has access to the intermediate outputs of models without knowledge of specific model parameters.
For the given image-text data point $(\mathbf{x}, \mathbf{c})$, we assume that $\mathbf{x}$ and $\mathbf{c}$ always correspond within the dataset $\mathbb{D}$. This assumption is evident in scenarios where dataset copyright owners perform membership inference to audit usage. And we also consider a weaker assumption of conducting membership inference without the groundtruth text in Sec.~\ref{sec: weak_assumption}.

Conversely, challenger $\mathcal{C}$ can mitigate the effectiveness of membership inference during training by utilizing data augmentation or even adaptive defense methods, which we discuss in Sec.~\ref{sec:defense}. Our work primarily focuses on fine-tuning scenarios because the weights of pretrained models are readily available, making this scenario more prone to copyright risks \cite{wang2023diagnosis, pang2023black}. Numerous projects are implemented by fine-tuning open-source models on specific datasets~\cite{text-to-image-ft, FT-CLOOB-Conditioned, Heart-of-Apple-XL, yeh2023navigating}.
We also conduct experiments on pretrained text-to-image diffusion models (Tab.~\ref{tab:wrap_pretrain}) to demonstrate the effectiveness of our method even in pretraining scenarios.

\subsection{Conditional Overfitting Phenomenon}\label{sec:key_i}
The rationale behind previous studies primarily hinges on the overfitting of diffusion models to training data (usually image data $\mathbf{x}$)~\cite{matsumoto2023membership, duan2023diffusion, kong2023efficient,chen2023robust,chen2024your}. This overfitting tends to result in lower estimation errors for images in the member set (training data) compared to those in the hold-out set during the diffusion process. Various indicators~\cite{matsumoto2023membership, duan2023diffusion, kong2023efficient} are designed based on this to expose membership information.
Specifically, let $q_{\text{mem}}$ and $q_{\text{out}}$ represent the image distributions of the member set and the hold-out set, respectively.  \( p \) represents the diffusion models' estimated distribution, and $D$ denotes a distance metric (which will be specified later). This rationale can be formulated as:
\begin{equation}
    D(q_{\text{mem}}(\vect{x}), p(\vect{x}))  \leq 
    D(q_{\text{out}}(\vect{x}), p(\vect{x})).
\end{equation}
However, if considering the membership inference on text-to-image diffusion models with image-text data \((\vect{x}, \mathbf{c})\), we emphasize the following assumption:
\begin{assumption}[Conditional overfitting phenomenon] \label{theorem:assumption}
The overfitting of text-to-image diffusion models to the conditional distribution of \((\vect{x}, \mathbf{c})\) is more salient than to the marginal distribution of \(\vect{x}\):
\begin{equation}\label{eq:overfit_conditional}
     \underbrace{\mathbb{E}_{\mathbf{c}}[D(q_{\text{out}}(\vect{x}|\mathbf{c}), p(\vect{x}|\mathbf{c})) - D(q_{\text{mem}}(\vect{x}|\mathbf{c}), p(\vect{x}|\mathbf{c}))]}_{\text{overfitting to conditional distribution}}  
       \geq \underbrace{D(q_{\text{out}}(\vect{x}), p(\vect{x})) - D(q_{\text{mem}}(\vect{x}), p(\vect{x})) }_{\text{overfitting to marginal distribution}}.
\end{equation}
\end{assumption}
\begin{wrapfigure}[15]{r}{0.53\textwidth}
\vspace{-0.4cm}
\centering
\includegraphics[scale=0.26]{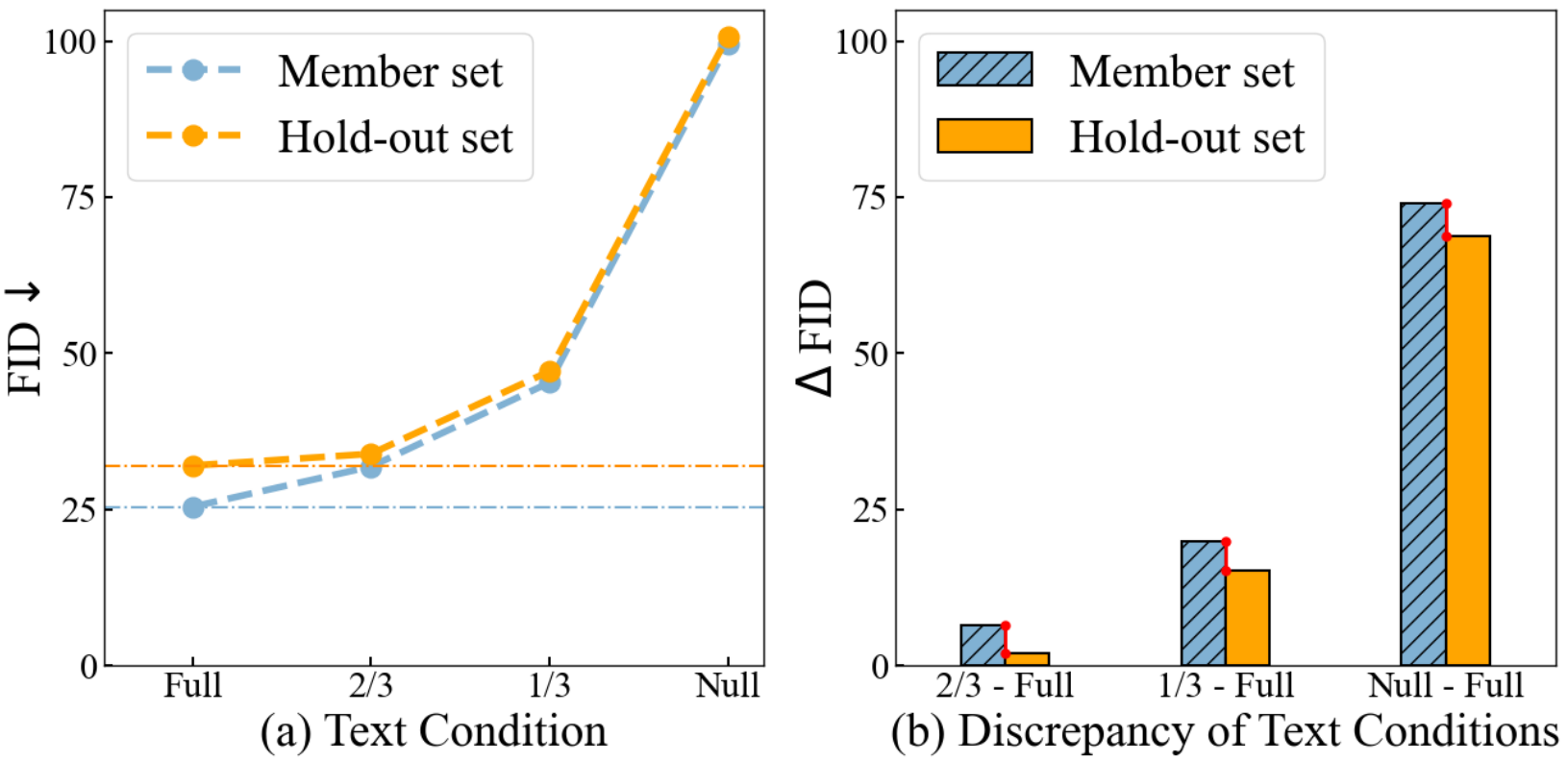}
\caption{
FID values   and the FID differences
of synthetic images ($2500/2500$ samples for member/hold-out set) under different conditions of member set and hold-out set.
}
\label{fig:intuition}
\end{wrapfigure}
Empirically, we validate this assumption by using Fréchet Inception Distance (FID)~\cite{heusel2017gans} as the metric $D$, i.e., \(D_{FID}\). 
We calculate $D_{FID}(q(\vect{x}|\mathbf{c}), p(\vect{x}|\mathbf{c}))$ using the  MS-COCO \cite{lin2014microsoft} dataset on a fine-tuned Stable Diffusion \cite{rombach2022high} model. Then by gradually truncating the original condition text to $\{2\big/3, 1\big/3, Null\}$ to obtain $\mathbf{c}^*$, we calculate $D_{FID}(q(\vect{x}|\mathbf{c}^*), p(\vect{x}|\mathbf{c}^*))$ as a stepwise approximation of  \(D_{FID}(q(\vect{x}), p(\vect{x}))\).
In Fig.~\ref{fig:intuition}, we report the FID scores of synthetic images under different conditions of member set and hold-out set. A smaller FID value indicates a closer match between model distributions and dataset distributions. 
From Fig.~\ref{fig:intuition} (a), it can be observed that for the full condition, the FID difference between the member set and the hold-out set is consistently higher than that for the truncated conditions, which validates our assumptions.  
We also demonstrate the validation utilizing other metrics in Appendix~\ref{appd:validation}.

We further compute the change in FID after truncating the condition and observe that the change in FID of the member set is consistently greater than that of the hold-out set (Fig.~\ref{fig:intuition} (b)), which inspires revealing membership 
by this overfitting discrepancy.
Recalling the aim of text-to-image diffusion model is to fit a latent space mapping from text to image, 
image data augmentation is commonly used to enhance the model generalization.
For instance, the official fine-tuning script of Hugging-Face~\cite{text-to-image-ft} employs Random-Crop and Random-Flip as the default augmentation \cite{text-to-image-ft-python}.
However, few trainers disturb the text condition as it is discrete and such disturbance would result in a decline of model utility (\cref{sec:defense}). 
Therefore, we believe that leveraging this phenomenon contributes to addressing the challenges of the strong generalization of text-to-image diffusion models with the resistance to data augmentation.

\subsection{Condition Likelihood Discrepancy}\label{sec:clid}
In this section, we derive a membership inference indicator for a given individual sample based on Assumption~\ref{theorem:assumption}.
Calculating FID requires sampling lots of images from the $p$ distribution, 
which is impractical under membership inference scenarios.
Instead, we employ Kullback-Leibler (KL) divergence as the distance metric, which is widely used and computationally convenient 
(the usage of other metrics is discussed in Appendix~\ref{appd:other_metric}). Then we have the following theorem:

\begin{theorem}
\label{theorem:theorem}
(Proof in Appendix~\ref{appd: proof})
When using \(D=D_{KL}\) as distance metric, Assumption~\ref{theorem:assumption} is equivalent to:
\begin{equation}
\label{eq:theorem}
       \mathbb{E}_{q_{\text{mem}}(\vect{x}, \vect{c})}[\log p(\vect{x}|\vect{c}) - \log p(\vect{x})]
          \geq        
       \mathbb{E}_{q_{\text{out}}(\vect{x}, \vect{c})}[\log p(\vect{x}|\vect{c}) - \log p(\vect{x})]
       + \delta_{H},
\end{equation}
where
\begin{equation}
    \delta_H = H(q_{\text{out}}(\vect{x})) + \mathbb{E}_{\vect{c}}[H(q_{\text{mem}}(\vect{x}|\vect{c}))]
       - H(q_{\text{mem}}(\vect{x})) - \mathbb{E}_{\vect{c}}[H(q_{\text{out}}(\vect{x}|\vect{c}))].
\end{equation}
\end{theorem}


Let us define: 
\begin{equation}\label{eq:indicator}
    \mathbb{I}(\mathbf{x},\mathbf{c}) = \log p(\vect{x} | \mathbf{c}) - \log p(\vect{x}).
\end{equation} 
If \(\delta_H\) is negligible, then according to \cref{eq:theorem}, it holds that \(\mathbb{E}_{q_{\text{mem}}}[\mathbb{I}(\mathbf{x})] \geq \tau \geq \mathbb{E}_{q_{\text{out}}}[\mathbb{I}(\mathbf{x})]\), where \(\tau\) is a constant intermediate between the left-hand side and right-hand side. 
Membership inference is then posed as follows: given an input instance \((\vect{x}, \mathbf{c})\), measuring \(\mathbb{I}(\mathbf{x}, \mathbf{c})\) to predict how probable it is that the input is a sample from \(q_{\text{mem}}\) rather than \(q_{\text{out}}\). 
Intuitively, if \(\mathbb{I}(\mathbf{x})\) exceeds a threshold \(\tau\), the instance is likely from \(q_{\text{mem}}\); otherwise, it belongs to \(q_{\text{out}}\). In the community of membership inference methods 
\cite{yeom2018privacy,chen2020gan,carlini2022membership, carlini2023extracting,matsumoto2023membership, duan2023diffusion, kong2023efficient, fu2023probabilistic}
, setting such a threshold \(\tau\) is a standard practice to differentiate between the two distributions. 
Therefore, we can utilize the indicator \(\mathbb{I}(\mathbf{x}, \mathbf{c})\) for membership inference.
Since Eq.~\eqref{eq:indicator}  actually involves measuring the likelihood discrepancy under different conditions of diffusion models, we call it \textbf{C}onditional \textbf{Li}kelihood \textbf{D}iscrepancy (\textbf{CLiD}).

In order to calculate the likelihoods in Eq.~\eqref{eq:indicator} for a given data point $(\mathbf{x},\mathbf{c})$, we utilize the ELBOs in Eq.~\eqref{eq:ddpm_elbo} and Eq.~\eqref{eq:cdm_elbo}  as an approximation of the log-likelihoods:
\begin{equation}\label{eq:CLiD_loss_ori}
     \mathbb{I}(\mathbf{x},\mathbf{c}) =
     \mathbb{E}_{ t, \mathbf{\epsilon}} \left[|| \mathbf{\epsilon}_{\theta}( \mathbf{x}_t    , t, \mathbf{c}_{\text{null}}) - \epsilon ||^2
     \right] 
     -  
     \mathbb{E}_{ t, \mathbf{\epsilon}} \left[
     || \mathbf{\epsilon}_{\theta}( \mathbf{x}_t    , t, \mathbf{c}) - \epsilon ||^2 \right],
\end{equation}
where $\mathbf{c}_{\text{null}}$ denotes an empty text condition input used to estimate the approximation of $\log p_\theta(\mathbf{x})$.

\subsection{Implementation of CLiD-MI}\label{sec:clid_mi}

In practice, calculating Eq.~\eqref{eq:CLiD_loss_ori} needs a Monte Carlo estimate for data point by 
sampling $N$ times using $(t_i, \epsilon_i)$
pairs, with $\epsilon_i \sim \mathcal{N}(\mathbf{0},\mathbf{I})$ and  $t_i \sim [1,1000] $. Performing two Monte Carlo estimations independently incurs high computational costs, resulting in $2 \times N$ query count, where  $N$ is typically a large number to ensure accurate estimation. To simplify computation, we instead perform Monte Carlo estimation on the difference of the ELBOs inspired by \cite{li2023your}:
\begin{equation}\label{eq:CLiD_loss}
    \mathbb{I}(\mathbf{x},\mathbf{c}) =
     \mathbb{E}_{ t, \mathbf{\epsilon}} \left[|| \mathbf{\epsilon}_{\theta}( \mathbf{x}_t    , t, \mathbf{c}_{\text{null}}) - \epsilon ||^2
     -  
     || \mathbf{\epsilon}_{\theta}( \mathbf{x}_t    , t, \mathbf{c}) - \epsilon ||^2 \right].
\end{equation}
In experiments, to further mitigate randomness, we also consider diverse reduced conditions along with  $\mathbf{c}_{\text{null}}$, forming the reduced condition set  
 $\mathbb{C} = \{\mathbf{c}^*_1, \mathbf{c}^*_2..., \mathbf{c}^*_{k}\}$, where we set 
 $\mathbf{c}^*_{k} = \mathbf{c}_{\text{null}}$. Then we compute multiple condition likelihood discrepancies:
\begin{equation}\label{eq:clid_loss_reduction}
    \mathcal{D}_{\mathbf{x},\mathbf{c},\mathbf{c}^*_i} = 
    \mathbb{E}_{ t, \mathbf{\epsilon}} \left[|| \mathbf{\epsilon}_{\theta}( \mathbf{x}_t    , t, \mathbf{c}^*_i) - \epsilon ||^2
     -  
     || \mathbf{\epsilon}_{\theta}( \mathbf{x}_t    , t, \mathbf{c}) - \epsilon ||^2 \right],
\end{equation}
where $\mathbf{c}_i^* \in \mathbb{C} $. In subsequent parts, we employ their mean or treat them as feature vectors to reveal membership information. We will introduce how to obtain $\mathbb{C}$ in \cref{sec:prac}.

\textbf{Combining $p_\theta(\vect{x}|\vect{c})$ for further enhancement.}
Recall that the practical significance of sample likelihood is the probability that a data point originates from the model distribution, which essentially can also be used to assess membership.
Due to the monotonicity of the log function, we can also use ELBO of Eq.~\eqref{eq:cdm_elbo} to estimate $p_\theta(\vect{x}|\vect{c})$:
\begin{equation}\label{eq:elbo_loss}
    \mathcal{L}_{\mathbf{x},\mathbf{c}} = - \mathbb{E}_{ t, \mathbf{\epsilon}} \left[|| \mathbf{\epsilon}_{\theta}( \mathbf{x}_t    , t, \mathbf{c}) - \epsilon ||^2 \right].
\end{equation}
Additionally, this estimation can reuse results from estimating Eq.~\eqref{eq:clid_loss_reduction}, thus obviating any additional query counts. Next, we consider two strategies to combine Eq.~\eqref{eq:clid_loss_reduction} and Eq.~\eqref{eq:elbo_loss} to construct the final membership inference method.

\textbf{Threshold-based attack--\clidth.}
First, we normalize the two indicators to the same feature scale. Due to the outliers in the data, we use Robust-Scaler: $\mathcal{S}(a_{i}) = (a_{i} - \tilde{a}) \big/ IQR $, where $a_{i}$ denotes the $i$-th value,  $\tilde{a}$ denotes the mean and IQR (interquartile range) is defined as the difference between the third quartile (Q3) and the first quartile (Q1) of the feature. Then we have:
\begin{equation}\label{eq:mi_th}
     \mathcal{M}_{\text{CLiD}_{th}}(\mathbf{x},\mathbf{c}) = \mathbbm{1} \left[
     \alpha \cdot \mathcal{S}(\frac{1}{k} \sum_{i}^{k}  \mathcal{D}_{\mathbf{x},\mathbf{c},\mathbf{c}^*_i})
     +
 (1-\alpha) \cdot \mathcal{S}(\mathcal{L}_{\mathbf{x},\mathbf{c}}) 
 > \tau  \right],
\end{equation}
where $k$ denotes the total number of reduced $\mathbf{c}^*$ (i.e., $k =  \left| \mathbb{C} \right|$),  and $\alpha$ is a weight parameter.

\textbf{Vector-based attack--\clidvec.}
We combine the estimated values of Eq.~\eqref{eq:clid_loss_reduction} and Eq.~\eqref{eq:elbo_loss} to obtain the feature vectors corresponding to each data point:
\begin{equation}\label{eq:vec}
    \mathbf{V} = 
\begin{pmatrix}
\mathcal{D}_{\mathbf{x},\mathbf{c},\mathbf{c}^*_1},
\mathcal{D}_{\mathbf{x},\mathbf{c},\mathbf{c}^*_2} 
\hdots
\mathcal{D}_{\mathbf{x},\mathbf{c},\mathbf{c}^*_k},
\mathcal{L}_{\mathbf{x},\mathbf{c}}
\end{pmatrix}.
\end{equation}
We use a simple classifier to distinguish feature vectors in order to determine the membership of the samples: 
\begin{equation}\label{eq:mi_vec}
\mathcal{M}_{\text{CLiD}_{vec}}(\mathbf{x},\mathbf{c}) = \mathbbm{1} \left[
\mathcal{F}_{\mathcal{M}} 
(
\mathbf{V}
)
> \tau  \right],
\end{equation}
where $\mathcal{F}_{\mathcal{M}}$ denotes the predict confidence of the classifier.

\subsection{Practical Considerations}\label{sec:prac}

\textbf{Reducing conditions to obtain $\mathbf{c}^*$.} 
We consider three methods for diverse reduction:
(1) Simply taking the first, middle, and last thirds of the sentences as text inputs. (2) Randomly adding Gaussian noises with various scales to the text embeddings. (3) Calculating the importance of words in the text  \cite{tang2023daam, wang2024promptcharm} and replacing them with “pad” tokens by varying proportions in descending order. For all three methods, we additionally use the null text input as $\mathbf{c}^*_k$. These methods are all effective and we use (3) with $k=4$ in subsequent experiments (details in Appendix~\ref{appd:detail}).

\textbf{Monte Carlo sampling.}  \label{sec:mtcl_sample}
 Let $M$ and $N$ denote the Monte Carlo sampling numbers of estimating $\mathcal{L}(\mathbf{x},\mathbf{c})$ and $\mathcal{D}_{\mathbf{x},\mathbf{c},\mathbf{c}^*_i}$, respectively.
 We set $M=N$ to achieve result reuse between Eq.~\eqref{eq:clid_loss_reduction} and Eq.~\eqref{eq:elbo_loss}, reducing the number of Monte Carlo sampling.
 Hence the overall query count of one data point is $M + K \cdot N$. Significant effects can be observed even when $M, N = 1$ (Fig.~\ref{fig:auc_samples}).

\textbf{Classifiers of \clidvec.} 
 Due to the simplicity of the feature vectors, we do not need a neural network as the classifier \cite{duan2023diffusion}. Simpler classifiers help to prevent overfitting.
In our experiments, we utilize XGBoost~\cite{chen2016xgboost} and utilize its predict confidence.

\section{Experiments}

\subsection{Setups}\label{sec:setup}

\textbf{Datasets and models.} For the fine-tuning setting,
we select $416 / 417$ samples on Pokémon~\cite{Pokemon-Blip}, $2500 / 2500$ samples on MS-COCO \cite{lin2014microsoft} and $10,000 / 10,000$ samples on Flickr \cite{young2014image} as the member/hold-out dataset, respectively. These three datasets involve diverse data distributions and dataset scales.
We use the most widely used text-to-image diffusion model, Stable Diffusion v1-4\footnote{ \url{https://huggingface.co/CompVis/stable-diffusion-v1-4}}~\cite{Stable-Diffusion-v1-4}, as the target model to fine-tune it on these three datasets. 
For the pretraining setting,
we conduct experiments on Stable Diffusion v1-5\footnote{\url{https://huggingface.co/runwayml/stable-diffusion-v1-5}}~\cite{Stable-Diffusion-v1-5} using the processed LAION dataset~\cite{schuhmann2022laion} (detailed in Sec.~\ref{sec:main_result}) to minimize distribution shift~\cite{dubinski2024towards,das2024blind}.

\textbf{Fine-tuning setups.}
For fine-tuning, previous membership inference on text-to-image diffusion models usually relies on strong overfitting settings. To evaluate the performance more realistically, we consider the two following setups: (1) \textit{Over-training.} Following the previous works~\cite{duan2023diffusion, kong2023efficient, fu2023probabilistic}, we fine-tune 15,000 steps on Pokemon datasets, and 150,000 steps on MS-COCO and Flickr (with only $2500 / 2500$ dataset size). (2) \textit{Real-world training.}  Considering that trainers typically do not train for such high steps, we recalibrate the steps based on the 
training steps/dataset size ratio (approximately 20) of official fine-tuning scripts on Huggingface~\cite{text-to-image-ft}. Thus, we train 7,500 steps, 50,000 steps and 200,000 steps for the Pokémon, MS-COCO and Flickr datasets, respectively. Additionally, we employ the default data augmentation (Random-Crop and Random-Flip~\cite{text-to-image-ft-python}) in training codes \cite{text-to-image-ft-python} to simulate real-world scenarios.

\textbf{Baselines.} We broadly consider existing member inference methods applicable to text-to-image diffusion models as our baselines: Loss-based inference~\cite{matsumoto2023membership}, SecMI$ _\text{stats} $ (SecMI)~\cite{duan2023diffusion}, PIA~\cite{kong2023efficient},  PFAMI$ _\text{Met} $ (PFAMI)~\cite{fu2023probabilistic} and an additional method of directly conducting Monte Carlo estimation (M. C.) on Eq.~\eqref{eq:elbo_loss} for comparison. For all baselines, we use the parameters recommended in their papers.
We omit some membership inference methods for generative models \cite{liu2019performing, chen2020gan, hayes2019logan}, as they have been shown ineffective for diffusion models in previous works \cite{duan2023diffusion, fu2023probabilistic}.  

\textbf{Evaluation metrics.}  
We follow the widely used metrics of previous works \cite{carlini2022membership,carlini2023extracting, duan2023diffusion, kong2023efficient, fu2023probabilistic}, including ASR (i.e., the accuracy of membership inference), AUC and the True Positive Rate (TPR) when the False Positive Rate (FPR) is 1\% (i.e.,  TPR@1\%FPR). 

\textbf{Implementation details.}
Our evaluation follows the setup of representative membership inference works  \cite{carlini2022membership, carlini2023extracting}. 
It is important to note that some implementations \cite{sec_implement, pia_implement} of previous works assume access to a portion of the exact member set and the hold-out set to obtain a threshold for calculating ASR or to train a classification network \cite{sec_implement}. This assumption does not align with real-world scenarios. 
Therefore, we strictly adhere to the fundamental assumption of membership inference \cite{carlini2022membership, fu2023probabilistic}: knowing only the overall dataset without any knowledge of the member/hold-out split. Hence, we first train a shadow model to obtain the 
$\alpha$ for Eq.~\eqref{eq:mi_th}, classifiers for Eq.~\eqref{eq:mi_vec}  and the threshold $\tau$ for 
calculating ASR with auxiliary datasets of the same distribution. Then we perform the test on the target models.
Other implementation details are provided in Appendix~\ref{appd:detail}.

\vspace{-7pt}
\subsection{Main Results}
\vspace{-3pt}

\label{sec:main_result}
\begin{table}[tbp]
  \centering
  \caption{Results under \textit{Over-training} setting. We mark the best and second-best results for each metric in \textbf{bold} and \underline{underline}, respectively. Additionally, the best results from baselines are marked in \blue{blue} for comparison.}
  \footnotesize
  \resizebox{\linewidth}{!}
  {
       \begin{threeparttable}
 \begin{tabular}{l ccc ccc ccc c}
    \toprule
    \multirow{2}[2]{*}{Method} & \multicolumn{3}{c}{MS-COCO} & \multicolumn{3}{c}{Flickr} & \multicolumn{3}{c}{Pokemon} & \multirow{2}[2]{*}{Query} \\
    \cmidrule(lr){2-4} 
    \cmidrule(lr){5-7}
    \cmidrule(lr){8-10}
          & ASR   & AUC   & TPR@1\%FPR   & ASR   & AUC   & TPR@1\%FPR & ASR   & AUC   & TPR@1\%FPR &  \\
    \midrule
    Loss
    & 81.92 & 89.98	& 32.28
    &81.90	&90.34	&40.80
   &83.76	&91.79	&25.77 
    & 1 \\
    PIA   & 68.56 & 75.12 & 5.08  
    &   68.56    &  75.12     &   5.08    
    & 83.37 & 90.95 & 13.31 
    & 2 \\
    M. C. 
    & 82.04 & 89.77 & 36.04
    & 83.32 & 91.37 & 41.20 
    & 79.35 & 86.78 & 23.74
    & 3 \\
    SecMI 
     &  83.00     & 90.81      &     50.64
    & 62.96$^\dagger$ & 89.29 & 48.52
    & 80.49 & 90.64 & 9.36  
    & 12 \\
    
     PFAMI 
    &\blue{94.48}	&\blue{98.60}	&\blue{78.00}
    &\blue{90.64}	&\blue{96.78}	&\blue{50.96}
    &\blue{89.86}	&\blue{95.70}	&\blue{65.35}
    & 20 \\
    \midrule
    \clidth & \underline{99.08} & \textbf{99.94} & \textbf{99.12} 
    & \underline{91.42} & \underline{97.39} & \textbf{74.00} 
    & \textbf{97.96} & \underline{99.28} & \textbf{97.84} & 15 \\
    \clidvec & \textbf{99.74} & \underline{99.31} & \underline{95.20} 
    &\textbf{91.78} & \textbf{97.52} & \underline{73.88} 
    & \underline{97.36} & \textbf{99.46} & \underline{96.88} & 15 \\
     \bottomrule
    \end{tabular}%
    \begin{tablenotes}
        \item $^\dagger$ When conducting SecMI~\cite{duan2023diffusion},  we observe that the thresholds obtained on the shadow model sometimes do not transfer well to the target model.
    \end{tablenotes}
    \end{threeparttable}
    \label{table:results_over}
    }
\end{table}%

\begin{table}[tbp]
    \vspace{-7pt}
  \centering
  \caption{Results under \textit{Real-world training} setting. We also highlight key results according to Tab.~\ref{table:results_over}.}
  \footnotesize
  \resizebox{\linewidth}{!}
  {
    \begin{tabular}{l ccc ccc ccc c}
    \toprule
    \multirow{2}[2]{*}{Method} & \multicolumn{3}{c}{MS-COCO} & \multicolumn{3}{c}{Flickr} & \multicolumn{3}{c}{Pokemon} & \multirow{2}[2]{*}{Query} \\
    \cmidrule(lr){2-4} 
    \cmidrule(lr){5-7}
    \cmidrule(lr){8-10}
          & ASR & AUC   & TPR@1\%FPR & ASR   & AUC   & TPR@1\%FPR & ASR   & AUC   & TPR@1\%FPR &  \\
    \midrule
    Loss
   & 56.28 &	61.89&	1.92
    &54.91	&56.60	&1.83
    &61.03	&65.96	&2.82 
    & 1 \\
     PIA   & 54.10 & 55.52 & 1.76  & 51.96 & 52.73 & 1.28  & 58.34 & 59.95 & 2.64  & 2 \\
    M. C. & 57.98 & 61.97 & 2.64  & 54.92 & 56.78 & 2.16  & 61.10 & \blue{66.48} & 3.84  
    & 3 \\
    SecMI & \blue{60.94} & \blue{65.40} & \blue{3.92}
    & \blue{55.60} & \blue{63.85} & \blue{2.76}  
    & \blue{61.28} & 65.56 & 0.84  
    & 12 \\
   
     PFAMI
    &57.36	&60.39	&2.72
    &54.68	&56.13	&1.80
    &58.94	&63.53	&\blue{5.76}
    & 20 \\
    \midrule
    \clidth 
    & \underline{88.88} & \underline{96.13} & \textbf{67.52} 
    & \underline{87.12} & \underline{94.74} & \underline{53.56} & \textbf{86.79} 
    & \textbf{93.28} & \textbf{61.39} 
    & 15 \\
    \clidvec 
    & \textbf{89.52} & \textbf{96.30} & \underline{66.36} 
    & \textbf{88.86} & \textbf{95.33} & \textbf{53.92} 
    & \underline{85.47} & \underline{92.61} & \underline{59.95} 
    & 15 \\
    \bottomrule
    \end{tabular}%
    \label{table:results_real}
    }
    \vspace{-9pt}
\end{table}%

\textbf{Over-training setting (fine-tuning).}
In Tab.~\ref{table:results_over}, models are trained for excessive steps on all three datasets, resulting in significant overfitting. We observe that under this over-training scenario, both of our methods nearly achieve ideal binary classification effectiveness. For instance, \clidth achieves over 99\% ASR, AUC and TPR@1\%FPR value on the MS-COCO dataset \cite{lin2014microsoft}.
With this training setup, the metrics for different baselines are very similar.  
Even the simplest loss-based method \cite{matsumoto2023membership} (with the query count of 1) also yields satisfactory results compared with other high query count methods.
Therefore, we emphasize: \textit{This unrealistic over-training setting fails to adequately reflect the effectiveness differences among various membership inference methods}.

\textbf{Real-world training setting (fine-tuning).}
In Tab.~\ref{table:results_real}, we adjust the training steps simulating real-word training scenario \cite{text-to-image-ft} and utilize default data augmentation \cite{text-to-image-ft-python}. The best value of ASR and AUC of baseline methods decreases to around 65\%, and the best value of TPR@1\%FPR decreases to around 5\%, indicating insufficient effectiveness of previous member inference methods in real-world training scenarios of text-to-image diffusion models. 
In contrast, our methods maintain ASR above 86\% and AUC above 93\%,  exceeding the best baseline values by about 30\%. The TPR@1\%FPR of our methods exceeds the best baseline values by  50\%\textasciitilde60\%.
The results demonstrate the effectiveness of our methods across various data distributions and scales in real-world training scenarios.

\begin{figure}[t]
\makeatletter\def\@captype{table}\makeatother
\begin{minipage}{0.45\linewidth}
\centering
\footnotesize
\resizebox{0.99\linewidth}{!} {
    \setlength{\tabcolsep}{5pt}
    \begin{tabular}{lccccccc}
    \toprule
    \multirow{2}[2]{*}{Method} &  \multicolumn{3}{c}{LAION} & \multirow{2}[2]{*}{Query} \\
  \cmidrule(lr){2-4}       & ASR   & AUC   & TPR@1\%FPR &  \\
    \midrule
    Loss      & 51.78 &	50.90 &	1.75
    & 1 \\
    PIA         & 52.13 & 52.42 & 1.25  & 2 \\
    M. C.        & 53.18 & 53.96 & 1.25  & 3 \\
    SecMI        & 57.43 & 58.59 & 2.45  & 12 \\
    PFAMI       & 59.08 & 61.11 &	1.45
    & 20 \\
    \midrule
    \clidth   & \textbf{64.53} & \textbf{67.82} & \textbf{5.01}  & 15 \\
    \bottomrule
    \end{tabular}%
}
\caption{The performance of membership inference methods on Stable Diffusion v1-5~\cite{Stable-Diffusion-v1-5} in pretraining setting. 
We utilize the processed LAION dataset to ensure the distribution consistency between member / hold-out sets~\cite{dubinski2024towards,das2024blind}.
The best results are highlighted in \textbf{bold}.}
  \label{tab:wrap_pretrain}%
\end{minipage}
\hspace{1ex}
\makeatletter\def\@captype{figure}\makeatother
\begin{minipage}{0.5\linewidth}
\centering
\vspace{-1em}
\includegraphics[scale=0.25]{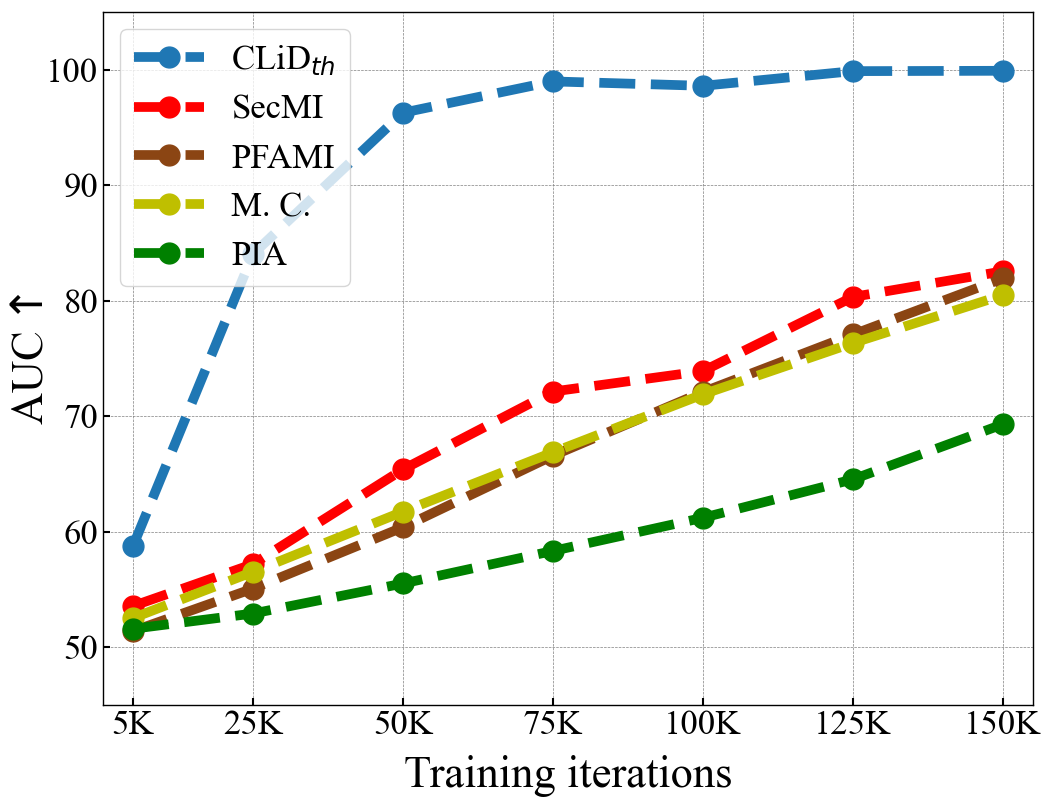}
\vspace{-0.5em}
\caption{Effectiveness trajectory on various training steps. 
}
\label{fig:auc_steps}
\end{minipage}
\vspace{-9pt}
\end{figure}

\textbf{Pretraining setting.}
For the pretraining setting, we adopt a stringent and realistic membership inference setting based on previous works~\cite{dubinski2024towards,das2024blind}. (1) To ensure the distribution consistency between the member and hold-out set, we respectively select $2500$ samples from the LAION-Aesthetics v2 5+ and LAION-2B MultiTranslated \cite{schuhmann2022laion} as member/hold-out set following~\cite{dubinski2024towards}; (2) We filter out samples containing non-English characters to ensure there are no other "distinguishable tails"~\cite{das2024blind} in the dataset\footnote{\citet{das2024blind} indicates that MultiTranslated-LAION dataset contains fewer non-English characters than the LAION dataset due to the use of the translation model. }.
We conduct membership inference on Stable Diffusion v1-5~\cite{rombach2022high}. 
As shown in Tab.~\ref{tab:wrap_pretrain}, our method consistently outperforms the baselines across all three metrics. 

\vspace{-2pt}
\subsection{Performance on Various Training Steps}
\vspace{-1pt}

\label{sec:vary_steps}
From Tab.~\ref{table:results_over} and Tab.~\ref{table:results_real}, we find that the training steps greatly influence the effectiveness of membership inference. 
All membership inference methods tend to exhibit satisfactory performance when the model is trained for an excessive number of steps that conflicts with real-world scenarios.
Therefore, we emphasize that the \textit{\textbf{effectiveness trajectory}} of membership inference across varying training steps should also be utilized to evaluate different methods. Better membership inference methods should reveal membership information earlier as training progresses.

To explore this, we fine-tune Stable Diffusion models with the MS-COCO dataset for varying training steps under \textit{real-world training} setting and report the AUC values of different membership inference methods in Fig.~\ref{fig:auc_steps}. 
It can be observed that as the training progresses, \clidth exhibits a significantly faster increase in effectiveness trajectory. By $25,000$ steps, \clidth effectively exposes membership information, whereas other baselines achieve similar results only at around $150,000$ steps. This demonstrates that our method can effectively reveal membership information when the overfitting degree of the text-to-image diffusion model is much weaker.

\vspace{-2pt}
\subsection{Ablation Study} 
\vspace{-1pt}

\label{sec:ablation}

\begin{wrapfigure}[17]{r}{0.45\textwidth}
\centering
\vspace{-1.5em}
\includegraphics[scale=0.24]{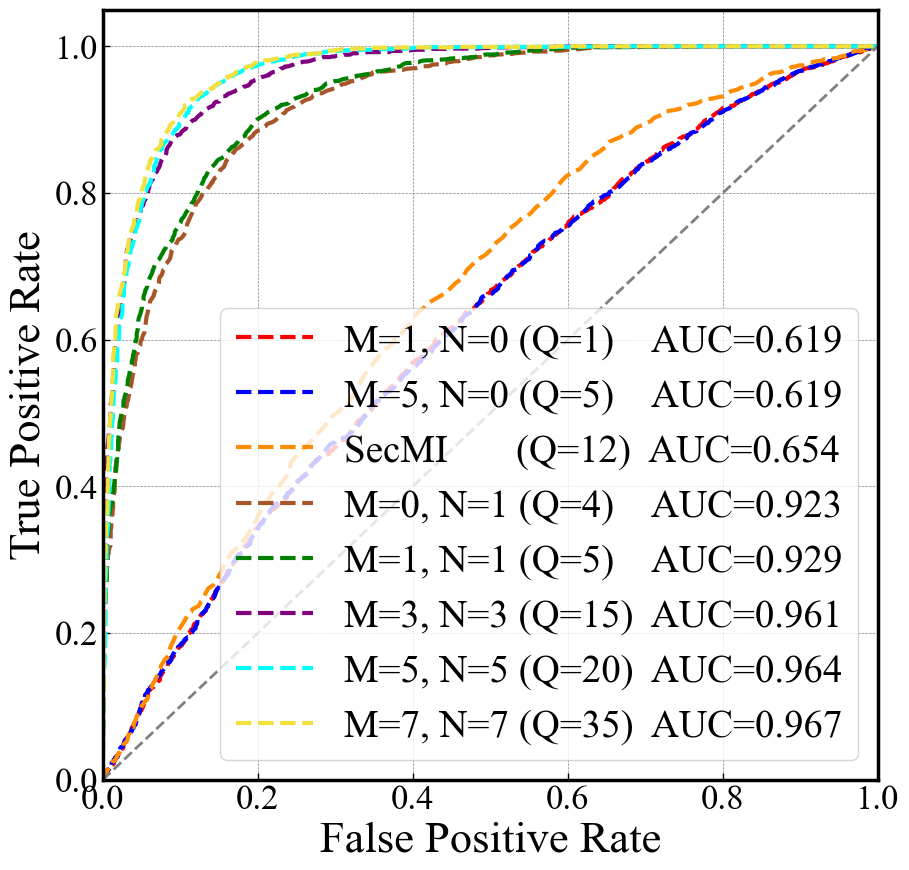}
\vspace{-1em}
\caption{Performance of \clidth and SecMI under various Monte Carlo sampling numbers (i.e., query count). 
The legend labels are sorted in ascending order by AUC values.
}
\label{fig:auc_samples}
\end{wrapfigure}
To conduct an ablation study, we vary the Monte Carlo sampling count in  Eq.~\eqref{eq:clid_loss_reduction}  and Eq.~\eqref{eq:elbo_loss}, perform $\text{CLiD}_{th}$ with MS-COCO dataset under \textit{real-world training} setting and report the AUC values in Fig.~\ref{fig:auc_samples}.   
To further compare the effects of Eq.~\eqref{eq:clid_loss_reduction}  and Eq.~\eqref{eq:elbo_loss}, we discard each term in Eq.~\eqref{eq:mi_th} and denote it as $M / N=0$.
We also include the result of the best baseline, SecMI~\cite{duan2023diffusion}, as a comparison.

\textbf{Effect of $ \mathcal{D}_{\mathbf{x},\mathbf{c},\mathbf{c}^*}$.} 
In Fig.~\ref{fig:auc_samples}, results of "M=1, N=0" and "M=1, N=1"  show a significant improvement of membership inference by including $ \mathcal{D}_{\mathbf{x},\mathbf{c},\mathbf{c}^*}$. Results of "M=5, N=0" and "M=1, N=1" further show that the method utilizing $\mathcal{D}_{\mathbf{x},\mathbf{c},\mathbf{c}^*}$ performs much better under the same sampling numbers. Additionally, the results of "M=0, N=1" and "M=1, N=1" indicates that only considering both Eq.~\eqref{eq:clid_loss_reduction} 
and Eq.~\eqref{eq:elbo_loss}  achieves the optimal performance.

\textbf{Monte Carlo sampling numbers.} 
In Fig.~\ref{fig:auc_samples}, we observe that when setting $M=N$, the performance improves as the number of Monte Carlo sampling increases. And the performance is improved slightly when $M, N>3$. Hence, we set $M, N=3$ 
to ensure the balance between a low query count and satisfied performance. Moreover, the experiment results of "M=1, N=1" and "SecMI" also demonstrate: \clidth outperforms previous works even with a much fewer query count.

\begin{table}[t]
\setlength{\tabcolsep}{8pt}
    \centering
  \caption{The performance of different methods under no augmentation and default augmentation. }
  \footnotesize
  \resizebox{\linewidth}{!}
  {
    \begin{threeparttable}
    \begin{tabular}{lcccccc}
    \toprule
    \multirow{2}[2]{*}{Method} 
          & \multicolumn{3}{c}{No Augmentation } & \multicolumn{3}{c}{Defaut Augmentation } \\
\cmidrule(lr){2-4}  \cmidrule(l){5-7}        & \multicolumn{1}{c}{ASR} & \multicolumn{1}{c}{AUC} & \multicolumn{1}{c}{TPR@1\%FPR} & \multicolumn{1}{c}{ASR ($\Delta$)} & \multicolumn{1}{c}{AUC ($\Delta$)} & \multicolumn{1}{l}{TPR@1\%FPR ($\Delta$)} \\
    \midrule
    Loss  
    & 66.54 &	72.73& 	7.72 
    &   56.28 (-10.26) &	61.89 (-10.84)& 	1.92 (-5.80) 
\\
 \textcolor{gray}{PIA}$^\dagger$  & \textcolor{gray}{56.56} &	\textcolor{gray}{59.28} &	\textcolor{gray}{2.00}
       &  \textcolor{gray}{54.10~(-2.46)} &	\textcolor{gray}{55.52~(-3.76)} &	\textcolor{gray}{1.76~(-0.24)}
 \\
    SecMI &   72.02	&81.07	&13.72
    &    60.94 (-11.08)   &   65.40 (-15.08)    &3.92 (-9.80)  \\
   
     PFAMI  &  79.20 & 	87.05& 	18.44 
      &  57.36 (-21.84)     & 60.39 (-26.66)      &2.72 (-15.72)  \\
    \midrule
    \clidth &  96.76 &	99.47 &	91.72
       &   88.88 \textbf{(-7.88)}    & 96.13 \textbf{(-3.34)}      &67.52 (-24.20)$^\ddagger$ \\
    \bottomrule
    \end{tabular}%
\begin{tablenotes}
    \item 
    $^\dagger$We omit the discussion of PIA as it shows no effectiveness at this training steps, with the metrics consistently approximating random guessing.
    \vspace{0.1cm}
    \item
    $^\ddagger$The TPR@1\%FPR value changes significantly here because its ROC curve is very sharp when FPR close to $0$.
    \end{tablenotes}
    \end{threeparttable}
    \label{tab:aug_defense}
    }
    \vspace{-1em}
\end{table}%

\begin{wraptable}{r}{0.45\textwidth}
\setlength{\tabcolsep}{2.5pt}
  \centering
  \vspace{-1.75em}
   \caption{Effectiveness of \clidth in adaptive defense. We calculate the FID \cite{heusel2017gans} with $10,000$ unseen MS-COCO samples to assess the model utility.}
   \vspace{0.5em}
      \resizebox{\linewidth}{!}
  {
      \begin{threeparttable}
   \begin{tabular}{lcccc}
    \toprule
    \multicolumn{1}{c}{\multirow{2}[2]{*}{Defense}} & \multicolumn{3}{c}{\clidth on MS-COCO} &  \multirow{2}[2]{*}{ FID $\downarrow$ / \textcolor{blue}{$\Delta$}} \\
\cmidrule{2-4}          & ASR   & AUC   & TPR@1\%FPR   \\
    \midrule
    None &88.88 	&96.13 &	67.52 
   & 13.17  \\
    \midrule
    Reph & 85.32 &	93.83 &	55.67      
    & 13.58 / \textcolor{blue}{+0.41} \\
    Del-1 & 86.40 &	93.59 &	59.52 
    & 13.18 / \textcolor{blue}{-0.01} \\
    Del-3 & 83.91 &	91.52 &	52.03 
    & 12.92 / \textcolor{blue}{-0.25} \\
    Shuffle & \textcolor{gray}{65.89} & \textcolor{gray}{67.37} & \textcolor{gray}{0.15}  & \textcolor{gray}{18.26 /  \textcolor{red}{+5.09}}$^\dagger$ \\
    \bottomrule
    \end{tabular}%
    \begin{tablenotes}
    \item
    $^\dagger$Compared to other methods, the increase in FID caused by shuffling is unacceptable for generative models.
    \end{tablenotes}
    \end{threeparttable}
  \label{tab:adaptive}%
  }
    \vspace{-0.4em}
\end{wraptable}
\vspace{-3pt}
\subsection{Resistance to Defense}\label{sec:defense}
\vspace{-3pt}

Since data augmentation is commonly used in training and can mitigate the effectiveness of membership inference~\cite{duan2023diffusion}, we use it to evaluate the performance 
 of methods under defense. 
As the baseline methods already exhibit weak performance under \textit{real-world training} setting, we opt not to incorporate additional data augmentation. Instead, we remove the default data augmentation from training scripts \cite{text-to-image-ft-python} to observe the effectiveness change of different methods. We fine-tune Stable Diffusion models for 50,000 steps with MS-COCO, report the metrics, and calculate the metrics changes in Tab.~\ref{tab:aug_defense}. 
We observe that the effectiveness of all membership inference methods declines after data augmentation is introduced during training. 
Note that PFAMI~\cite{fu2023probabilistic} exhibits the highest sensitivity to data augmentation since it infers membership by probability fluctuation after images are perturbed, which also explains its significant performance decline between Tab.~\ref{table:results_over} and Tab.~\ref{table:results_real}.
Compared to the baselines, our method exhibits the smallest decrease, which indicates its stronger resistance to data augmentation. 

\textbf{Adaptive defense.} 
We further consider adaptive defense: assuming the trainers are aware of our methods and perturb the text of image-text datasets before training. We consider the following adaptive defense methods: (1) rephrasing the original text\footnote{We utilize ChatGPT-3.5 with the following prompt: "Please rewrite the following sentences while keeping the key semantics."}, (2) randomly deleting 10\%, 30\% words in text, and (3) shuffling 50\% of the image-text mappings in the dataset.
In Tab.~\ref{tab:adaptive}, we observe that except for \textit{shuffling}, the other adaptive defense methods have almost no effect on \clidth. And \textit{shuffling} damages the model utility (too high FID values), rendering this defense meaningless.

\vspace{-3pt}
\subsection{Weaker Assumption}\label{sec: weak_assumption}
\vspace{-3pt}
Although in Sec.~\ref{sec:threat_m} we assume that the adversary can access the entire image-text pairs based on the real-world data usage auditing scenario, we also consider a weaker assumption: the adversary can only access the image without the corresponding text.

In this scenario, we first generate pseudo-text corresponding to the images using an image captioning model (BLIP~\cite{li2022blip} in our experiments), and then conduct CLiD-MI based on the image-pseudo\_text pairs. In Tab.~\ref{table:weak_assum}, we observe that our method still broadly outperforms baselines. We believe this is because the pseudo-text preserves the image's key semantics, keeping our methods effective.


\begin{table}[t]
    \setlength{\tabcolsep}{11pt}
\centering
  \caption{Results without access to the corresponding text 
  under \textit{Over-training} setting and  \textit{Real-world training} setting. We fine-tune MS-COCO on SDv1-4. 
  Key results are highlighted as Tab.~\ref{table:results_over}.
  }
  \footnotesize
  \resizebox{\linewidth}{!}
  {
    \begin{tabular}{l ccc ccc c}
    \toprule
    \multirow{2}[2]{*}{Method} & \multicolumn{3}{c}{\textit{Over-training} (Pseudo-Text)} & \multicolumn{3}{c}{\textit{Real-world training} (Pseudo-Text)} & \multirow{2}[2]{*}{Query} \\
    \cmidrule(lr){2-4} 
    \cmidrule(lr){5-7}
          & ASR & AUC   & TPR@1\%FPR & ASR   & AUC   & TPR@1\%FPR  &  \\
    \midrule
  Loss  & 73.80 & 81.01 & 9.71  & 56.08 & 58.47 & 1.60  & 1 \\
    PIA   & 61.40 & 65.75 & 1.20  & 53.44 & 54.38 & 1.52  & 2 \\
    M. C. & 74.36 & 81.55 & 11.28 & 56.68 & 60.00 & 1.28  & 3 \\
    SecMI & 82.04 & 88.97 & 40.80 & \blue{60.48} & \blue{64.04} & \blue{3.28} & 12 \\
    PFAMI & \blue{91.56} & \blue{95.16} & \blue{68.16} & 58.12 & 59.77 & 2.64  & 20 \\
    \midrule
    \clidth & \underline{92.84} & \underline{95.43} & \textbf{72.36} & \underline{76.16} & \underline{83.27} & \textbf{19.76} & 15 \\
    \clidvec & \textbf{93.26} & \textbf{96.59} & \underline{71.73} & \textbf{77.76} & \textbf{84.48} & \underline{18.06} & 15 \\
    
    \bottomrule
    \end{tabular}%
    \label{table:weak_assum}
    }
    \vspace{-5pt}
\end{table}%

\section{Related Works}
\vspace{-3pt}

\textbf{Copyright protection in text-to-image synthesis.} 
To protect the copyright of text-to-image models, several works~\cite{zhao2023recipe,zhai2023text} propose inserting backdoors to embed watermarks in text-to-image models.
To protect the copyright of image-text datasets, some works~\cite{salman2023raising, zhao2023unlearnable, shan2023glaze} incorporate imperceptible perturbations to render the released datasets unusable.
Other works~\cite{wang2023diagnosis, cui2023ft} utilize the backdoor or watermark to track the usage of image-text datasets.  
In contrast, our method indicates the possibility of auditing the unauthorized usage of individual image-text data points utilizing membership inference.

\textbf{Membership inference on diffusion models.} 
In the grey-box or white-box setting, \citet{carlini2023extracting} firstly conduct membership inference on unconditional diffusion models by conducting LiRA (Likelihood Ratio Attack) \cite{carlini2022membership}, with the requirement of training multiple shadow models. \citet{matsumoto2023membership} make the first step by utilizing diffusion loss to conduct query-based membership inference. Some works~\cite{duan2023diffusion, kong2023efficient} leverage the DDIM~\cite{song2020denoising} deterministic forward process~\cite{kim2022diffusionclip} to access the posterior estimation errors of diffusion models. And \citet{fu2023probabilistic} leverage the probability fluctuations by perturbing image samples. Few works consider the black-box settings \cite{wu2022membership, pang2023black}. However, these studies either assume partial knowledge of member set data \cite{wu2022membership} or assume extensive fine-tuning steps~\cite{pang2023black} ($100 \sim 500$ epochs), both of which do not align with real-world scenarios.

\textbf{Memorization detection in text-to-image models.}
A similar work~\cite{wen2024detecting} 
detects token memorization by inspecting the magnitude of text-conditional predictions, 
but differs from ours by lacking in-depth rationale analysis and a rigorous membership inference setup with randomly selected member/hold-out sets.

\vspace{-3pt}
\section{Conclusion}\label{sec:conclusion}
\vspace{-3pt}
In this paper, we identify the phenomenon of conditional overfitting in text-to-image models and propose \textbf{CLiD-MI}, the membership inference framework on text-to-image diffusion models utilizing the derived indicator, conditional likelihood discrepancy. Experimental results demonstrate the superiority of our method and its resistance against early stopping and data augmentation. Our method aims to inspire a new direction for the community regarding unauthorized usage auditing.

\textbf{Limitations:} 
Due to the limited availability of open-source text-to-image diffusion models, evaluations under the pretraining setting are not sufficient. Considering fine-tuning setting involves a multi step/image ratio, 
we acknowledge that the superiority of \textbf{CLiD-MI} over the baselines in the pretraining setting  is not as evident compared to fine-tuning setting.
We emphasize our experiments under pretraining setting (Tab.~\ref{tab:wrap_pretrain}) reveal the hallucination success of existing works and encourage future research to focus on this more challenging and practical scenario.

\vspace{-5pt}
\section*{Acknowledgments}
\vspace{-5pt}
We thank anonymous reviewers for their valuable feedback. In addition, we thank Xin Zhang for his editorial comments.
This work is supported by the National Key R\&D Program of China (No.2022YFB2703301), NSFC Projects (Nos.~92370124, 62076147). Y. Dong is also supported by the China National Postdoctoral Program for Innovative Talents and Shuimu Tsinghua Scholar Program.


\clearpage
{
\small
\bibliographystyle{plainnat}
\bibliography{main}
}
\clearpage


\appendix
\numberwithin{equation}{section}
\numberwithin{figure}{section}
\numberwithin{table}{section}
\setcounter{theorem}{0}

\section{Validation of Assumption~\ref{theorem:assumption} }
\label{appd:validation}

\begin{figure}[h]
    \centering
    \includegraphics[width=\textwidth]{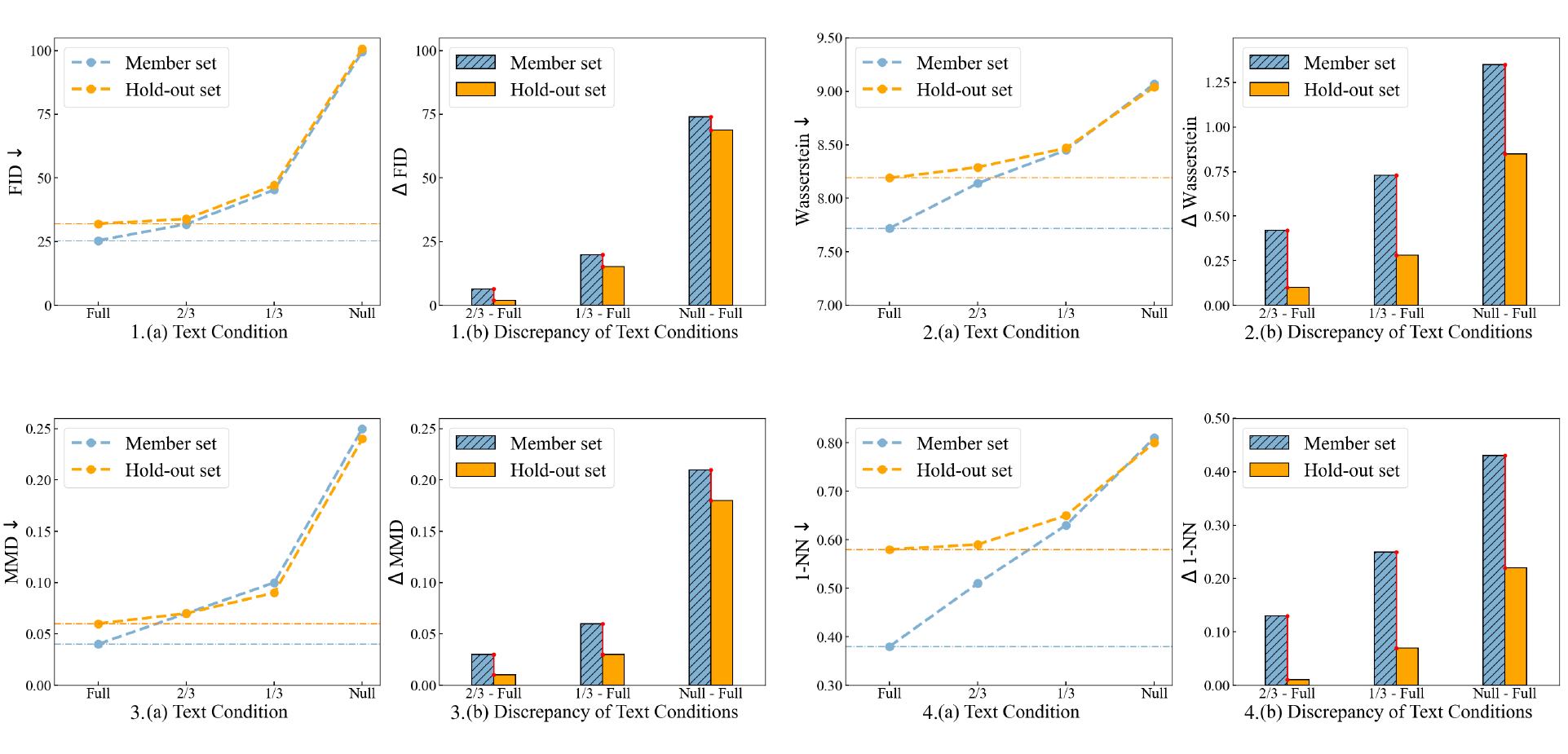}
    \caption{Metric values and the metric differences of synthetic images, with the same setting as Sec.~\ref{sec:key_i}. 
    }
    \label{fig:metric4appd}
\end{figure}

To extensively validate the effectiveness of Assumption~\ref{theorem:assumption}, we utilize additional metrics as the distances metric $D$ in Eq.~\eqref{eq:overfit_conditional}, including \textit{Wasserstein Distance}~\cite{xu2018empirical}, \textit{Kernel MMD (Maximum Mean Discrepancy)}~\cite{xu2018empirical} and  \textit{1-Nearest Neighbor Classifier (1-NN)}~\cite{lopez2016revisiting}, in addition to FID~\cite{heusel2017gans}. As observed in Fig.~\ref{fig:metric4appd}, regardless of the metric used for $D$, Assumption~\ref{theorem:assumption} consistently holds, thereby confirming the broad applicability of \textbf{Conditional Overfitting} phenomenon.

\section{Proof of Theorem~\ref{theorem:theorem}}\label{appd: proof}
\begin{proof}
Eq.~\eqref{eq:overfit_conditional} is equivalent to:
\begin{equation}
   \begin{aligned}
       & \mathbb{E}_{\vect{c}}[D(q_{\text{out}}(\vect{x}|\vect{c}), p(\vect{x}|\vect{c}))] -  D(q_{\text{out}}(\vect{x}), p(\vect{x})) \\     \geq  
       &  \mathbb{E}_{\vect{c}}[D(q_{\text{mem}}(\vect{x}|\vect{c}), p(\vect{x}|\vect{c}))] - D(q_{\text{mem}}(\vect{x}), p(\vect{x})).
   \end{aligned}
\label{eq:overfit_conditional_change}
\end{equation}
Given that both the member set and the hold-out set are mixtures of Dirac distributions:
\begin{equation}
    q(\vect{x})=\frac{1}{|D_{set}|} \sum_{\vect{x}_i \in D_{set}} \delta(\vect{x}-\vect{x}_i),
\end{equation}
where \(D_{set}\) denotes the set of images in the corresponding dataset. We can derive the analytical form for the Kullback-Leibler (KL)  KL divergence when using \(D_{KL}\) as the distance metric:
\begin{equation}
    \begin{aligned}
         D_{KL}(q(\vect{x}), p(\vect{x})) 
         &=\int q(\vect{x}) \log \frac{q(\vect{x})}{p(\vect{x})}d\vect{x} \\
        &=- \int (\frac{1}{|D_{set}|} \sum_{\vect{x}_i \in D_{set}} \delta(\vect{x}-\vect{x}_i)) \log p(\vect{x})dx + H(q(\vect{x})) \\
        &=- \frac{1}{|D_{set}|} \sum_{\vect{x}_i \in D_{set}}  \int \delta(\vect{x}-\vect{x}_i) \log p(\vect{x})dx + H(q(\vect{x}))\\
        &=- \frac{1}{|D_{set}|} \sum_{x_i \in D_{set}}  \log p(x_i) + H(q(\vect{x})).\\
    \end{aligned}
\end{equation}
where  \(H\) is the entropy functional. Therefore, we have:
\begin{equation}\label{eq:substi_dirac}
    \begin{aligned}
        &\mathbb{E}_{\vect{c}}[D_{KL}(q(\vect{x}|\vect{c}), p(\vect{x}|\vect{c}))] -  D_{KL}(q(\vect{x}), p(\vect{x})) \\
        = &-\frac{1}{|D_{set}|} \sum_{\vect{x},\vect{c} \in D_{set}} [\log p(\vect{x}|\vect{c}) - \log p(\vect{x})] + \mathbb{E}_{\vect{c}}[H(q(\vect{x}|\vect{c}))]-H(q(\vect{x})),
    \end{aligned}
\end{equation}
where $D_{set}$ is the corresponding dataset (member set or hold-out set). Substituting Eq.~\eqref{eq:substi_dirac} into Eq.~\eqref{eq:overfit_conditional_change}, we can get:
\begin{equation}\label{eq:proof_result0}
    \begin{aligned}
      & -\frac{1}{|D_{\text{out}}|} \sum_{\vect{x},\vect{c} \in D_{\text{out}}} [\log p(\vect{x}|\vect{c}) - \log p(\vect{x})]
       + \mathbb{E}_{\vect{c}}[H(q_{\text{out}}(\vect{x}|\vect{c}))]-H(q_{\text{out}}(\vect{x}))
    \\ \geq  
       & -\frac{1}{|D_{\text{mem}}|} \sum_{\vect{x},\vect{c} \in D_{\text{mem}}} [\log p(\vect{x}|\vect{c}) - \log p(\vect{x})]
       + \mathbb{E}_{\vect{c}}[H(q_{\text{mem}}(\vect{x}|\vect{c}))]-H(q_{\text{mem}}(\vect{x})).
    \end{aligned}
\end{equation}
Eq.~\eqref{eq:proof_result0} is equivalent to:
\begin{equation}\label{eq:proof_result1}
    \begin{aligned}
      & -  \mathbb{E}_{q_{\text{out}}(\mathbf{x}, \vect{c})} [\log p(\vect{x}|\vect{c}) - \log p(\vect{x})]
       + \mathbb{E}_{\vect{c}}[H(q_{\text{out}}(\vect{x}|\vect{c}))]-H(q_{\text{out}}(\vect{x}))
    \\ \geq  
       & -  \mathbb{E}_{q_{\text{mem}}(\mathbf{x}, \vect{c})} [\log p(\vect{x}|\vect{c}) - \log p(\vect{x})]
       + \mathbb{E}_{\vect{c}}[H(q_{\text{mem}}(\vect{x}|\vect{c}))]-H(q_{\text{mem}}(\vect{x})).
    \end{aligned}
\end{equation}
Finally, we can get:
\begin{equation}
       \mathbb{E}_{q_{\text{mem}}(\vect{x}, \vect{c})}[\log p(\vect{x}|\vect{c}) - \log p(\vect{x})]
          \geq        
       \mathbb{E}_{q_{\text{out}}(\vect{x})}[\log p(\vect{x}|\vect{c}) - \log p(\vect{x})]
       + \delta_{H},
\end{equation}
where
\begin{equation}
    \delta_H = H(q_{\text{out}}(\vect{x})) + \mathbb{E}_{\vect{c}}[H(q_{\text{mem}}(\vect{x}|\vect{c}))]
       - H(q_{\text{mem}}(\vect{x})) - \mathbb{E}_{\vect{c}}[H(q_{\text{out}}(\vect{x}|\vect{c}))].
\end{equation}
\end{proof}

\section{Metrics Discussion of \cref{sec:clid} }\label{appd:other_metric}
In the derivation of \cref{sec:clid}, we also consider other metrics besides KL divergence. The results show that KL divergence yields the most easily computable analytical form. For instance, we briefly discuss Jensen–Shannon (JS) divergence as follows:

Recall the expression for Jensen-Shannon divergence:
\begin{equation}
    D_{JS}(q, p) = D_{KL}(q, \frac{1}{2}(q+ p)) + D_{KL}(p, \frac{1}{2}(q + p). 
\end{equation}
The first parameter of the KL divergence should be a simple distribution that is easy to compute; otherwise, deriving the analytical form for such divergence is typically difficult. 
In Eq~\eqref{eq:overfit_conditional}, JS divergence cannot be efficiently computed because it includes \(D_{KL}(p, \frac{1}{2}(q + p))\), where $p$ denotes the model distribution.
It needs to use the Monte Carlo method, which involves sampling images from both \(q\) and \(p\) to make an approximation. As a result,  this process is extremely time-consuming.

\section{Experiment Details}\label{appd:detail}

\subsection{Monte Carlo Sampling}

\begin{figure}[t]
\centering
\includegraphics[width=0.6\columnwidth]{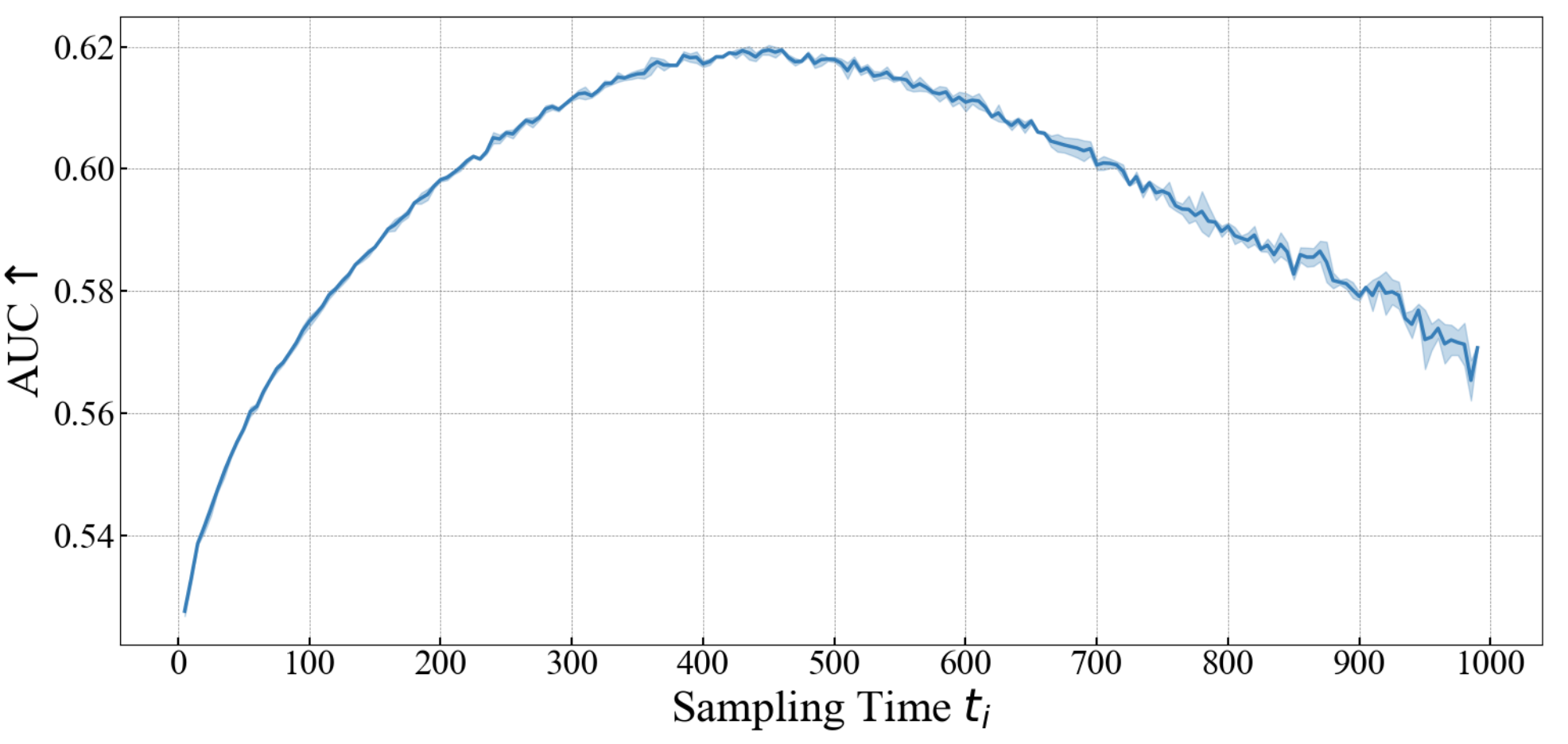}
\caption{Effectiveness of single  Monte Carlo estimation of various timesteps. Small $t_i$ corresponds to less noise added, and large $t_i$ corresponds to significant noise.  AUC value is highest when the timestep is around $450$.}
\label{fig:mtcl_1q}
\end{figure}

In our method, the key to accurate membership inference lies in estimating ELBO with fewer sampling steps for better precision.
To achieve this, firstly, we reduce the number of Monte Carlo samples by directly estimating the ELBO difference (Eq.~\eqref{eq:CLiD_loss}). Secondly, recalling Monte Carlo sampling using $(t_i, \epsilon_i)$ pairs with $\epsilon_i \sim \mathcal{N}(0,\mathbf{I})$ and  $t_i \sim [1,1000] $, we explore the effect of the sampling time $t_i$. 
We conduct a single Monte Carlo sampling test using MS-COCO on \textit{real-word training} setting and report the AUC values in Fig.~\ref{fig:mtcl_1q}.

In Fig.~\ref{fig:mtcl_1q}, we observe that the single Monte Carlo estimation achieves optimal accuracy when $t_i \in [400,500]$. Similar results are shown in \cite{li2023your}. Therefore, consistent with \cite{li2023your}, we sample at intervals of $10$ centered around the timestep $450$.
In our experiments,   
$M,N$ in Eq~\eqref{eq:clid_loss_reduction} are both uniformly set to $3$ (i.e., the estimation number
is $3$), and we use the time list of $[440,450,460]$, resulting in the query count of 15.
Note that \cite{carlini2023extracting,duan2023diffusion} indicate that for DDPM of Cifar-10~\cite{krizhevsky2009learning}, the best estimation timestep is around $100$. 
This difference may arise from the different signal-to-noise ratios of images with various resolution~\cite{hoogeboom2023simple}.
This finding suggests that the Monte Carlo sampling timestep should be designed differently for diffusion models of different scales.

\subsection{Reduction Methods}

In implementation, we actually diversely reduce the condition $\mathbf{c}$ to $\mathbf{c}^*$ and calculate $p_\theta(\mathbf{x} | \mathbf{c}^*)$ to approximate $p_\theta(\mathbf{x} )$.
In this part, we evaluate the effectiveness of different reduction methods. We consider three methods in Sec.~\ref{sec:prac}: (1) Simply Clipping. We simply use the first, middle, and last thirds of the sentences as text inputs. (2) Gaussian Noise. We add Gaussian noises with the scales of $50\%,70\%,90\%$ to the overall text embeddings. (3) Importance Clipping.  We calculate the importance of words in the text\footnote{\url{https://github.com/ma-labo/PromptCharm}}~\cite{tang2023daam, wang2024promptcharm} and replace them with “pad” tokens in descending order by varying proportions of $30\%,50\%,70\%$. For all three methods, we additionally use the null text as a $\mathbf{c}^*$. The experiments are conducted on the \textit{real-world training} setting with MS-COCO dataset. And we also employ null text solely to compute Eq.~\eqref{eq:clid_loss_reduction} without reduction methods for comparison. 

In Tab.~\ref{tab:reduction}, we observe that \textit{Importance Clipping} achieves the best results due to its more deterministic reduction. So we adopt it as the reduction method used in our experiments. Additionally, we note that all three reduction methods exhibit satisfactory results, demonstrating the general applicability of our method. Comparing the results without the usage of reduction methods, the results validate the effectiveness of reduction methods in Sec.~\ref{sec:prac}.

\begin{table}[t]
  \centering
  \caption{The membership inference performance with different reduction methods. "Null" denotes employing null text solely to compute Eq.~\eqref{eq:clid_loss_reduction} without reduction methods.}
    \begin{tabular}{lcccc}
    \toprule
    \multirow{2}[2]{*}{Reduction Methods} & \multicolumn{3}{c}{\clidth on MS-COCO} & \multirow{2}[2]{*}{Query} \\
\cmidrule{2-4}          & ASR   & AUC   & TPR@1\%FPR &  \\
    \midrule
    Null (K=1) & 85.10  & 93.60  & 42.96  & 6 \\
    Simply Clipping (K=4) & 88.02  & 95.90  & 66.53  & 15 \\
    Gaussian Noise (K=4) & 86.58 & 94.79 & 56.78 & 15 \\
    Importance Clipping (K=4) & \textbf{88.88} & \textbf{96.13}  & \textbf{67.52}  & 15 \\
    \bottomrule
    \end{tabular}%
\label{tab:reduction}
\end{table}%

\section{Compute Overhead and Resources}\label{appd:compute}
\textbf{Computational Overhead.} As a query-based member inference method, the computational efficiency of our method primarily depends on the number of queries. A lower query count signifies a more efficient member inference method. Our method significantly outperforms the baselines when the query count are about the same (such as SecMI and PFAMI in Sec.~\ref{sec:main_result}). Furthermore, even with a much lower query count such as $M=1, N=1 (Q=5)$ (Fig.~\ref{fig:auc_samples}), our method exhibits a noticeable improvement compared to the baselines.

\textbf{Compute Resources.} 
Our experiments are divided into two main parts: training (fine-tuning) and inference, both conducted on a single RTX A6000 GPU. The time of execution in the training phase depends on the training steps. For example, we perform $7,500$, $50,000$, and $200,000$ steps for Pokemon~\cite{Pokemon-Blip}, MS-COCO~\cite{lin2014microsoft} and Flickr~\cite{young2014image} dataset, which take about 2 hours, 12 hours, and 48 hours, respectively.
The time of execution in inference time depends on the methods' query count. For example, with the query count of $15$,  our membership inference method on a dataset of size $2500/2500$ takes approximately $80$ minutes per run for all data points. Typically, we perform this inference once on the shadow model and once on the target model, resulting in a total time cost of $160$ minutes.

\section{Ethics Statements}\label{appd:ethics}
Although the current threat models for membership inference methods include privacy attack scenarios and data auditing scenarios, we emphasize that for text-to-image diffusion models, the potential application of membership inference lies more in unauthorized data usage auditing than in data privacy leakage. This is because most training data is obtained by scraping open-source image-text pairs, which are more likely to pose copyright threats rather than privacy violations. So we emphasize that our method can make a positive societal impact for inspiring unauthorized usage auditing technologies of text-image datasets in the community.


\end{document}